\documentclass[prl,10pt,twocolumn,showpacs,amssymb,amsmath]{revtex4-1}
\usepackage[english]{babel}  
\usepackage{mathrsfs,amsfonts,graphics}

\def\bra#1{\mathinner{\langle{#1}|}}
\def\ket#1{\mathinner{|{#1}\rangle}}

\def\Bra#1{\left\langle#1\right|}
\def\Ket#1{\left|#1\right \rangle}
{\catcode`\|=\active 
  \gdef\Braket#1{\begingroup
\mathcode`\|32768\let|\BraVert\left<{#1}\right>\endgroup}}
\def\BraVert{\egroup\,\mid\,\bgroup}

\usepackage{color}
\definecolor{Blue}{rgb}{0,0,1}
\definecolor{Red}{rgb}{1,0,0}
\definecolor{Green}{rgb}{0,1,0}
\definecolor{Purp}{rgb}{.2,0,.2}
\definecolor{white}{rgb}{1,1,1}

\begin{document}

\title{Preparation of states in open quantum mechanics}

\author{Kavan Modi}
\email[Email: ]{kavan@quantumlah.org}
\affiliation{Centre for Quantum Technologies, National University of Singapore, Singapore}
\affiliation{Department of Physics, The University of Texas at Austin, Austin TX, USA}

\begin{abstract}
We study preparation of states for open quantum mechanics. For non-Markovian systems that are initially correlated with the environment, the affects of the preparation procedure are nontrivial. This is due to the indirect affects on the state of the environment induced via the correlations with the system and the act of preparation on the system. We give three concrete examples of preparation procedure to elucidate our claims.
\end{abstract}

\date{\today}
\maketitle

\emph{Introduction.}---Correlations have a fundamental role in quantum theory.  Quantum correlations like entanglement are credited for the speedup of quantum computation over classical computation, highly secured communication, enhancement of precision in metrology, etc. On the other hand, unwanted effects such as decoherence and dephasing are due to the correlations with the surrounding degrees of freedom. The theoretical studies of the former are dealt with quantum information theory \cite{Nielsen00a} and of the latter with the theory of open quantum systems \cite{SudarshanMatthewsRau61, kossakowski,*lindblad}.  Nowadays, the two disciplines are intimately related as the practical application of quantum information processors experience open dynamics.  While the theory of open quantum mechanics hopes to gain better foundations from the principles of quantum information theory. In this Letter we discuss a topic that concerns both of these disciplines at a fundamental level, state preparation for correlated quantum systems.

There is a  great deal of of literature regarding state preparation, however, much of the discussion revolves around preparing specific states for specific systems, e.g. squeeze states in optical systems, entangled states in solid-state qubits, etc. Yet, this topic has received little foundational attention since the early years of quantum theory, especially for open systems.  The lack of fundamental attention for theory of preparation was pointed out some years ago by Park and Band \cite{parkband}. They beautifully remark that a preparation is not to be confused with a measurement. They go on to elaborate on many different ways to prepare quantum states, including mixed states and point out the differences between different preparation procedures and discuss the usefulness of each.  

A quantum experiment can be thought of as composed of three steps. The experiment begins with an unknown state, which is prepared into a known state, called an input. The input is subjected to some quantum operation yielding an output. Finally, the state of the output is observed.  Mathematically this corresponds to the composition of three quantum operations as
\begin{equation}
\mathcal{M}_n \circ \mathcal{B} \circ \mathcal{A}_{m} (\rho)=o_{mn},
\end{equation} 
where $\mathcal{A}$ represents the action of the preparation apparatus for the $m$th input, $\mathcal{B}$ represents the dynamics, and $\mathcal{M}$ represents the $n$th measurement giving the output result $o_{mn}$. The preparation and the dynamics are though to be independent of each other. However this may not always be true. Several authors~\cite{PhysRevLett.75.3020, PhysRevA.64.062106, kuah:042113} have suggested that when dealing with open systems the preparation procedures play a non-trivial role in quantum process tomography experiments . In this Letter we extend those results by developing a general theory for preparation procedures for open quantum systems. We show how the act of preparation affects quantum systems that are initially (before preparation) correlated with the surrounding degrees of freedom. That is, it is not always possible to speak of the operations of dynamics and preparation as being independent.  In a mathematical sense we have $(\mathcal{B\circ A})_m\neq\mathcal{B}\circ\mathcal{A}_m$. The result of this inseparability of the two operations is that the act of preparation contributes to the `quantum memory' when the dynamics of the system is of non-Markovian nature.


\emph{Motivation.}---Let us first motivate this paper with a simple example. Consider a single qubit experiment that requires two pure states, say $\ket{x,+}$ and $\ket{y,+}$. Suppose the experimentalist has an apparatus that initiates qubits with no preferred polarization. The experimentalist has to then prepare this fully-mixed qubit into the desired state.  There are several ways of making the desired state, let us consider two. (\emph{i}) Subject the qubit to projective operation along the $x$-direction.  The qubit will collapse into one of two states: $\ket{x,+}$ or $\ket{x,-}$. Discard if the outcome is $\ket{x,-}$. Otherwise proceed with the experiment with $\ket{x,+}$. To prepare the $\ket{y,+}$ state, subject fully-mixed qubit to projective operation along the $y$-direction and follow the same prescriptions as before. (\emph{ii}) Prepare $\ket{x,+}$ state same as in the first procedure.  But instead of discarding the $\ket{x,-}$ beam, unitarily rotate it to state $\ket{y,+}$. The question that we tackle in this Letter is: \emph{are the two sets of preparations the same?} In turn, are the two experiments the same? Below we show that they indeed are not when the experiment is \emph{open} and the system is \emph{initially correlated} with the surrounding degrees of freedom.


\emph{Preparations.}---In practice, preparation procedures can be very complicated. Instead of describing many different preparation procedures in detail, we develop a general theory of preparations in terms of the standard mathematical tools as follows. A preparation procedure is the mapping of a set of unknown states, $\rho$, into a fix known input state, $P_{m}$. The most general transformation of a quantum state is described by a stochastic map \cite{SudarshanMatthewsRau61}. In light of that, a very complicated preparation procedure due to a complicated apparatus is simply denote by a stochastic map, $\mathcal{A}_{m}$.  The procedure of preparing the $m$th input state is given as
\begin{eqnarray}\label{prep1}
\rho\longrightarrow \frac{1}{a_{m}}\mathcal{A}_{m}(\rho)=P_{m},
\end{eqnarray}
where $a_{m}=\mbox{Tr}[\mathcal{A}_{m}(\rho)]$ is the probability with which $\rho$ goes to $P_{m}$.  The map $\mathcal{A}$ is required to be completely positive but not a trace preserving. This because a preparation procedure takes any state to a set of fixed states, therefore its domain must be set of all system states, hence it must be completely positive. While many preparation procedures are implemented strictly by getting rid of particles in unwanted states, e.g. procedure $(i)$ above.

A typical Experiments requires multiple input states and according to Eq. \ref{prep1} each input corresponds to different a preparation map, hence a different preparation procedure.  The main result of the paper is show how different preparation procedures affect the experiment differently for open quantum systems. For open quantum systems, the situation above does not change very much, one can simply replace the unknown state of the system by an unknown state of the system and the environment, transforming Eq. \ref{prep1} into
\begin{eqnarray}\label{prepopen}
\rho^{\mathcal{SE}}\longrightarrow
\left[\frac{1}{a_{m}}\mathcal{A}_{m}
\otimes\mathcal{I}\right]
\left(\rho^{\mathcal{SE}}\right)=R^\mathcal{SE}_{m}.
\end{eqnarray}
Above $\mathcal{I}$ is the identity map acting on the state of the environment~\footnote{It is not necessary to assume that the preparation procedure does not act on the environment. But, since we do not know what happens to the environment we can assume this and get by without further complicating the equations.}.  In other words, we are assuming that the preparation procedure acts only on the system and not the state of the environment.  However, any action on the system also means an action on the correlations that the system shares with the environment. In general, the state of the system and the environment before preparation, $\rho^\mathcal{SE}$, will be correlated (see~\cite{modigeo,Ferraro-2010a}). The goal of the preparation procedure is to eliminate these correlations, as for an ideal experiment the state of the system should be uncorrelated with the environment at the beginning of the experiment. Typically, preparation procedures achieve this by preparing pure states, which guarantees the product form for the system and the environment. In that case the right hand side of Eq. \ref{prepopen}, or the prepared state, becomes
\begin{eqnarray}\label{prepapx}
\frac{1}{a_{m}}\mathcal{A}_{m}
	\left(\rho^{\mathcal{SE}}\right)=P_{m}\otimes\rho^\mathcal{E}_{m},
\end{eqnarray}
where the subscript $m$ of the state of the environment in the last equation means that the state of the environment has dependency on the $m$th preparation procedure.


It should be mentioned that if the initial state of the system and the environment is uncorrelated, $\rho^{\mathcal{SE}}=\rho^{\mathcal{S}}\otimes \rho^{\mathcal{E}}$, then the act of preparation does not effect the environment. Furthermore, if the interaction between the system and the environment is Markovian, then no information from the environment flows back to the system \cite{lainePRL} and therefore the system does not care whether the state of the environment is dependent on the preparation procedure or not. The difference between preparation procedures for closed systems and open systems lie entirely in the correlations shared by the system and the environment. The preparation procedure acts only on the system, the state of environment is affected by that action due to the shared correlations.

Eq. \ref{prepapx} is the main result of this Letter.  It shows that the act of preparation affects the state of the environment. This is not a trivial issue for a system that experiences non-Markovian dynamics \cite{breuer02a, cesarnonmarkov}.  In such instances, the state of the environment acts as the `memory' of the system. Meaning, this information will come into the system at a latter time.  According to Eq. \ref{prepapx}, for different input states, the state of the environment will be different and therefore the corresponding memory term will be different as well. At this point, we would like to describe three preparation procedures.  The first procedure will clearly show this dependency of the state of the environment on the preparation procedure. The second procedure aims to eliminate this dependency.  And finally the last procedure shows how inconsistent implementation of the second procedure can lead to the state of the environment depending on the preparation procedure.


\emph{Projective Preparation.---}Consider an optical experiments, where a beam of photons is sent through a polarizer \cite{Nielsen:1998,PhysRevA.64.012314,PhysRevLett.91.120402}, which projects the polarization vector of the photons into a set of orthogonal directions. We call this \emph{projective preparation} method.  Mathematically this is summarized as
\begin{eqnarray}\label{mesprep}
\frac{1}{a_{m}}\mathcal{A}_{m}\left(\rho^{\mathcal{SE}}\right)
=\frac{1}{a_{m}} 
P_{m} \rho^\mathcal{SE} P_{m}
=P_{m}\otimes\rho^\mathcal{E}_{m},
\end{eqnarray}
where  $P_{m}$ is the $m$th projector in the system space as well as the desired input state and $a_{m} =\mbox{Tr}[\mathcal{A}_{m}\rho] =\mbox{Tr}[P_{m}\rho]$; this is the probability of obtaining that particular input state from a von Neumann measurement. The state of the environment has picked the subscript $m$ due to the fact that in general $\rho^\mathcal{SE}$ contains correlations between the system and the environment. We should again emphasize that the preparation procedure only acts on the system, the state of the environment is affected only indirectly. The analysis of projective preparation shows that the state of the environment after the preparation depends on the preparation procedure due to the pre-existing correlations with the system.  Should the environment and the system interact in non-Markovian fashion then the dynamics for different input states will differ due to the differences in the states of the environment.  The dependency of the environment on the preparation procedure can lead to unwanted behavior of the experiment \cite{kuah:042113,modidis}.  For this reason this dependency may be unwanted for experimental purposes. Note that this procedure is the same as the one discussed earlier in the scenario (i). Let us now discuss a method of preparation that keeps the state of the environment constant for any input state of the system.


\emph{Stochastic preparation.---}Many quantum experiments begin by initializing the system into a specific state. For instance, in the simplest case, the system can be prepared to the ground state by cooling it to near absolute zero temperature \cite{Wein:121.13, orien:080502, Howard06, myrskog:013615}. Mathematically, these set of operations are written as a pin map \cite{GoriniSudarshan},
\begin{eqnarray}
\Theta=\ket{\Phi}\bra{\Phi}\otimes\mathbb{I},
\end{eqnarray}
where $\mathbb{I}$ is the identity matrix, acting as the `trace operator' ($\mathbb{I}_{rs}\rho_{rs}=\mbox{Tr}[\rho]$) and $\ket{\Phi}$ is some fixed state (i.e. ground state) of the system. In this procedure, no matter what the initial state of the system was, it is ``pinned" to the final state $\ket{\Phi}\bra{\Phi}$. The action of the pin map $\Theta$ on a bipartite state of the system and the environment.
\begin{eqnarray}\label{pinmap}
\Theta \left(\rho^\mathcal{SE}\right)
&=&\left[\Ket{\Phi}\Bra{\Phi}\otimes\mathbb{I} \right]
\left(\rho^\mathcal{SE}\right)
=\Ket{\Phi}\Bra{\Phi}\otimes\rho^\mathcal{E}.
\end{eqnarray}
The pin map fixes the system into a single pure state, which means that the state of the environment is fixed into a single state as well; the pin map decorrelates the system from the environment. Once the pin map, $\Theta$, is applied, the system is prepared in the various different input states by applying local operations
\begin{eqnarray}\label{stocmap}
\mathcal{A}_{m}\left(\rho^\mathcal{SE}\right)
&=&\nonumber
\Omega_{m} \circ\Theta(\rho^\mathcal{SE})
= P_{m}\otimes\rho^\mathcal{E}.
\end{eqnarray}
As seen the last equations, the advantage of stochastic preparation method, as described here, is that the state of the environment is a constant for any input of the system.  This constancy of the state of the environment is  necessary to characterize the dynamical mechanisms properly \cite{modidis}.  This procedure corresponds to scenario (ii) discussed earlier. Looking at Eqs. \ref{mesprep} and \ref{stocmap}, it is easy to see now that the two scenarios are fundamentally different from each other as the states of the environment are different.

A word of caution is necessary at this point.  It may seem that stochastic preparation procedure alleviates the experiment from having an environment that depends on the preparation procedure. We analyze this matter in much greater detail elsewhere \cite{prep2}, however, for the last example we would like discuss how the procedure above can go wrong when the stochastic method is implemented in an inconsistent manner.


\emph{Multiple Stochastic Preparations.---}Strictly speaking, in a real experiment the pin map is a complicated set of steps.  For instance, cooling the system so it relaxes to the ground state is very different from optically pumping a system to an excited state.  Two different pin maps would correspond to these two different methods.  Then, the state of the environment in each case can also be different.  We assumed, in Eq. \ref{prepopen}, that the preparation procedure does not act on the environment.  We may generalize Eq. \ref{prepapx} by considering that each preparation procedure also effects the state of the environment in a direct manner:
\begin{eqnarray}\label{prepapx}
\frac{1}{a_{m}}\mathcal{A}_{m}\otimes\mathcal{Q}_m
	\left(\rho^{\mathcal{SE}}\right)=P_{m}\otimes\mathcal{Q}_m(\rho^\mathcal{E}),
\end{eqnarray}
where $\mathcal{Q}_m$ is an operation on the state of the environment due to the $m$th preparation procedure. In such instance, preparation procedure becomes important even if there were no initial correlation between the system and the environment. This easily seen in the case of applying multiple stochastic-preparation procedures for preparing different input states.

If a stochastic preparation procedure acts directly on the environment, to be careful, we should label $\rho^\mathcal{E}$ as $\mathcal{Q}(\rho^\mathcal{E})$ to clarify that the state of the environment may depend on certain steps involved in applying the pin map. It may be tempting to simply use a set of pin maps, $\Theta_{m}$, to prepare the various input states $P_{m}$.  The different input states will be
\begin{eqnarray}
\Theta_{m}\otimes\mathcal{Q}_m\left(\rho^\mathcal{SE}\right)
= P_{m}\otimes\mathcal{Q}_m(\rho^{\mathcal{E}})
= P_{m}\otimes\rho^{\mathcal{E}}_m.
\end{eqnarray}
As seen in the last line of the equation above, the state of the environment has taken the subscript $m$ through different pin maps along with the input state.  What this means is that the state of the environment is directly effected by the preparation procedure. In this situation, the effect remains even if there were no initial correlations between the system and the environment.

In \cite{prep2} several examples of quantum process tomography using the three preparation procedures stated above are given. There it is shown that for initially correlated system and environment states that have non-Markovian interactions, some preparation procedures fail to give a linear, completely-positive process map. A remedy to this is give in \cite{prep3}, where a quantum process tomography procedure is developed that yields a map that is independent of the preparation procedure.

\emph{Conclusions.}---What we have demonstrated in this Letter is that a preparation procedure can play a role for initially correlated states of the system and the environment. The affects of the preparation may show up in the dynamics of the system because they effect the state of the environment which in return affects the system. However, if the experiment is closed in nature, then there is no environment to worry about and any set preparation procedure are equivalent. If the system is initially uncorrelated with the surrounding, then generally the preparation procedure will not play a role in dynamics.  However, there will be some cases where this rule will not hold (see \cite{prep2} for an example). Three, if the interaction with the environment is of Markovian type, i.e. the environment does not kick back the system, then the preparation procedures will not affect the dynamics of the system. 

{\bf Acknowledgments.} We are grateful to Aik-Meng Kuah, Ali Rezakhani, C\'esar Rodr\'iguez-Rosario, George Sudarshan, and Daniel Terno for valuable conversations. This work was financially supported by the National Research Foundation and the Ministry of Education of Singapore.

\bibstyle{is-unsrt}
\bibliography{preparationQPT} 									
\end{document}